# Metamaterial atom with multijunction superconducting structure.

A.V. Shadrin*, S.E. Bankov, G.A. Ovsyannikov, K.Y. Constantinian.

Kotel'nikov Institute of Radio Engineering and Electronics, Russian Academy of Sciences,
Mokhovaya st. 11-7, 125009 Moscow, Russia.

*anton_sh@hitech.cplire.ru

We present results of numerical simulations of a metamaterial element (meta-atom) for planar negative media based on cuprate Josephson junctions. The proposed circuit consists of two coupled resonators: "magnetic dipole" and "electric dipole". The first one is a superconducting spiral resonator with multijunction superconducting structure - SQIF (SQIF - superconducting quantum interference filter), and the second one is a comb-shape resonator made from normal metal film. We show, that in comparison with passive metal resonators, the superconducting resonator with Josephson junctions (magnetic dipole) can be tuned by external magnetic field due to magnetic dependence of Josephson inductance. The simulated characteristics of one-dimensional planar negative media are discussed.

Keywords:
Meta material, negative media, tunable resonator, cuprate Josephson structure, superconducting quantum interference filter.

Introduction.

Recently theoretical and experimental studies of metamaterials based on linear passive reciprocal structures were reported (see [1, 2] and references therein). These studies identified the problems to be solved for realization of unique features of metamaterials: not sufficient small level of microwave loss and tunability. The superconducting circuits with low losses look promising, particularly ones based on superconducting Josephson junctions [3-6]. Recently it was demonstrated that superconducting spiral resonator could be used in metamaterial atom where each individual mode offers a narrow-band window of effective negative-permeability [7]. Unlike ordinary passive resonators [8] thin film resonator with a Josephson junction can be tuned under external magnetic field which changes the Josephson inductance. An artificial circuit with negative refraction factor at microwave frequencies requires a combination of electric and magnetic dipoles [9, 10]. Usually, electric dipoles are Hertzian dipoles, while the magnetic ones are split-rings with narrow cut-slot [8]. Following this approach we made modeling of the layout of the basic cell element of metamaterial (meta-atom) for one-dimensional negative media based on Josephson superconducting multijunction structure, constituting superconducting quantum interference filter (SQIF).

*1. Meta-atom model.*

To evaluate properties of the proposed meta-atom we use a model of a cell consisted of three transmission line sections (see inset in Fig. 1). Signal transmission microstrip line has characteristic impedance $Z_0$ and electric length $\theta_0 = 2\pi l/\lambda_0$, were $l$ is the length of microstrip line and $\lambda_0$ is wave length at the frequency $f_0$. The second section with impedance $Z_1$ and electric length $\theta_1$ at the frequency $f_1$ forms the electric resonator, directly coupled with the signal transmission line, represents the electric dipole. The third section with impedance $Z_2$ and electric length $\theta_2$ at frequency $f_2$ models the "magnetic dipole". Coupling between resonators is implemented by the capacitor $C_c$. The position of connecting point of capacitor $C_c$ could be shifted by $\Delta\theta_1$ in the electrical dipole and $\Delta\theta_2$ in the magnetic dipole. Resistors $R_{1,2}$ simulate dissipative loss. Without coupling between resonators, $C_c=0$, the input signal is almost totally reflected from the electric dipole at the resonance frequency $f_1$ and transmission coefficient tends to zero. Calculated frequency dependences of dispersion parameters $S_{11}$ and $S_{12}$ are shown in Fig. 1 for the case $C_c$=8 pF, $Z_2$=200 Ω, $\theta_2$=170º, $\Delta\theta_2$=10º, $R_2$=5·10$^5$ Ω и $f_2$=2.84 GHz. Coupling between dipoles results in appearance of a narrow transparency band. Its frequency position depends on resonators tuning and could be considered as an optimum one when both central frequencies of the transparency band and the stop band are close to each other. One can observe an increasing of meta-atom transparency at the resonant frequency of magnetic dipole. However, in the transparency band the transmission coefficient $S_{12}$ doesn't reach unity due to existence of loss (see curve 2 in Fig.1). The transmission coefficient $S_{12}$ depends on the loss in the magnetic dipole, while the losses in the electric dipole have not such strong effect. Thus, it is reasonable to use an extremely low loss superconducting magnetic dipole and an electric dipole made from nonsupercoducting material.

Calculations of $S_{12}$ for different $R_2$ show that transmission coefficient decreases with loss increase. Note that if resistance $R_2$ is increased over some level its further increase corresponds to loss decreasing. When $R_2 = \infty$ the resonator reaches the running wave regime and losses are reduced to minimum. The transmission bandwidth depends on the level of coupling between resonators: decreasing the coupling capacitance from $C_c$=8 pF to $C_c$ =5 pF the transmission bandwidth decreases nearly by three times.

*2. Meta-atom implementation using multijunction superconducting circuit.*

The meta-atom based on superconducting multijunction structure is shown in Fig.2. Here meta-atom consists of two weakly coupled resonators. "Magnetic dipole" is a superconducting spiral double line with a SQIF [11], and electric dipole is a comb shape resonator made from thin metal film. SQIF is placed in the central part of the double-line spiral resonator. We use serial SQIF where dc SQUIDs are connected in series (see Figure 3). The length of the spiral resonator was chosen corresponding to the half-wave resonance within frequency range $f$=1–3 GHz. Experimental SQIFs were fabricated using $NdGaO_3$ substrates which dielectric constant is $\varepsilon$=22. Critical currents (and self-inductances $L_J$) of Josephson junctions and SQUIDs are very sensitive to the external magnetic field [11, 12] and served as parameters for frequency tuning. The electric dipole with high Q-factor used for sufficiently well microwave coupling of superconducting spiral resonator with external signal.

Fig.3 shows the typical SQIF topology. The thin film circuit was formed by means of ion-plasma and chemical etching of $YBa_2Cu_3O_x$ film deposited onto $NdGaO_3$ bicrystal substrate using dc sputtering at high oxygen pressure. The SQIF structure consist of 30 serially connected SQUIDs with various superconducting loop areas in the range 50–400 $\mu m^2$. The width $w$ of Josephson junctions was 10 μm with tolerance no more than ±0.1 μm. Magnetic flux was determined by the field generated by external magnetic coil.

Fig.4 shows results of spectral characteristics of simulation carried out for frequency range $f$= 0–3 GHz when electric dipole is loaded by 50 Ω line. Single standing electric dipole shows a minimum around 2 GHz. Inserting a magnetic dipole we observe a change in response: narrow peaks appear at $f$=1.2 GHz (very weak, hence hardly seen well in the figure) and at $f$=2.013 GHz. Because of differences between the radiation losses for electric and magnetic dipoles their Q-factors differ significantly. In a system of two coupled resonators at $f$= 2.013 GHz we see an increase in the transparency for the whole structure. A similar behaviour was observed earlier for response from two split-rings, coupled with the electric dipole [13].

*3. Magnetic field control of multi-junction superconducting circuit*

When magnetic field is applied to the SQIF the resonance frequency of magnetic dipole must shift due to change of inductances of Josephson junctions. An important parameter for this meta-atom is the modulation depth of Josephson inductance under influence of external magnetic field. According to our simulations the resonance frequency of magnetic dipole varies with the inductance of Josephson junctions in the SQIF with a rate near to 2 MHz/pH (see Fig. 4).

Let us evaluate the dependence of SQIF inductance from external magnetic field. Josephson junction inductance is determined by critical current $I_C$ and phases difference $\varphi$:

$$L_J = \Phi_0 (2\pi I_C \cos\varphi)^{-1} \qquad (1)$$

where $\Phi_0 = h/2e$ is a quantum of magnetic flux. Total inductance of the $i^{th}$ SQUID in SQIF includes also the geometric inductance $L_{Ki}$. Assuming that parameters of Josephson junctions are identical, for a SQUID with two JJ connected in parallel the sum of inductances is $L_J/2 + L_{Ki}/2$. Accordingly, total inductance of SQIF with $N$ serially connected SQUIDs with different $L_{Ki}$ is as following:

$$L_\Sigma = 1/2 \sum_{i=1}^{N} (L_{Ki} + L_J) \qquad (2)$$

From (1) and (2) the change in inductance of the SQIF under the influence of external magnetic flux $\Phi_e$:

$$\frac{dL_\Sigma}{d\Phi_e} = \frac{N dL_J}{2 d\Phi_e} = \frac{N}{2}\left(-\frac{\hbar \frac{dI_C}{d\Phi_e}}{2eI_C^2 \cdot \cos\varphi} + \frac{\hbar \cdot \sin\varphi \cdot \frac{d\varphi}{d\Phi_e}}{2eI_C \cdot \cos^2\varphi}\right) \qquad (3)$$

Here the phase difference in JJ is changed by magnetic flux $\Phi_e$. Total magnetic flux $\Phi$ in the $i^{th}$ SQUID with the screening current $I_s$, induced by the external magnetic flux $\Phi_e$ is the sum of two terms [14]:

$$\Phi = \Phi_e + I_s \cdot L_{ki} \qquad (4)$$

For the case of large geometric inductances, $L_{Ki} \gg L_J$, that usually takes place in experiment for most of SQIF structures, the SQUIDs act as superconducting rings with very weak influence of JJs, conserving magnetic flux unchanged. In this case external flux $\Phi_e$ is compensated by current $I_s$, and the phases of JJs in SQUIDs become independent [10]. This allows us to neglect the second term in (3) and use simple Fraunhofer dependence for critical current of single JJ:

$$I_C = I_C^0 \frac{\sin \pi \frac{\Phi_e}{\Phi_0}}{\pi \frac{\Phi_e}{\Phi_0}} \qquad (5)$$

We see that in the SQIF with $N$ serially connected SQUIDs with large inductive loop inductances the variation $L_{Ki}$ is $dL_{\Sigma}/d\Phi_e$ is due to change in the critical current by magnetic field, $dI_C/d\Phi_e$. For a thin-film JJ and known magnetic field (and magnetic flux) we use well known relation $\Phi_e = \mu_0 H w^2$ [15], where $w$ is the width of bicrystal JJ. Taking $\cos\varphi \sim 1$ for the case when external flux $\Phi_e = \Phi_0/2$ the change of the critical current $I_C$ in (5), according to (1-3), is $dL_{\Sigma}/d\Phi_e = N/4I_C$. Thus, in the case of the SQIF structure with $N = 30$, $w = 10$ μm, and the typical value of critical current $I_C = 100$ μA we estimate $dL_{\Sigma}/d\mu_0 H = 7.5\ 10^{-6}$ H/T.

According to the measurements data for the SQIF with $N = 30$ SQUIDs [11, 12], we can change the slope of the critical current in SQIF with magnetic field $dI_C/d\mu_0 H = 100$ A/T. Thus, we obtain $dL_{\Sigma}/d\mu_0 H = \Phi_0/(2\pi I_C^2) dI_C/d\mu_0 H = 3\ 10^{-6}$ H/T, which is slightly less than the estimation made in our simulation. Then, again, for typical value of the critical current $I_C = 100$ μA obtain the frequency variation magnitude $df/d\mu_0 H = 3 \cdot 10^{-6} \cdot f_r/L_{\Sigma}$ Hz/T, where $f_r$ is the resonant frequency of SQIF -structure.

Thus, for realistic change $\mu_0 H = 100$ μT of magnetic field we obtain a shift of resonance frequency of SQIF about 600 MHz for $f_r = 2$ GHz and $L_{\Sigma} = 1$ nH. Here it is assumed that the Fraunhofer dependence (5) dominates in SQIF based on bicrystal JJs with relatively large superconducting loops. It's worth to note that thin film superconducting loops in SQIF increase sensitivity to external magnetic field even in the absence of flux transformer. Note also, in SQIF -structure with large geometric inductances, $L_{Ki} >> L_J$, the spread of critical currents is less critical, since the main contribution of inductance modulation comes from magnetic field dependence of the critical current of individual bicrystal JJs.

*4. 1D meta-medium based on multi-element superconducting structures.*

We see there is at least a qualitatively agreement between model parameters obtained for the resonant cells calculations and the results for electrodynamic simulations. It allows us to use such approach for synthesis of 1D metamaterial medium and evaluating it's parameters. 1D meta-medium could be formed as a periodic structure made by cascading combination of meta-atoms (see inset in Fig. 5).

Let's consider the eigenmode waves in such structure and choose a period with length $P$ for the cell characterized by scattering matrix $\hat{S}$. Amplitudes of incident waves $U_{1,2i}$ at the inputs 1 and 2 are connected with amplitudes of reflected waves $U_{1,2r}$ by elements of scattering matrix:

$$U_{1r} = U_{1i} S_{11} + U_{2i} S_{12}, \qquad (6)$$

$$U_{2r} = U_{1i}S_{21} + U_{2i}S_{22}.$$

Due to the cell symmetry and its reciprocity for parameters of scattering matrix, one may write the following relations:

$$S_{11} = S_{22}, \ S_{12} = S_{21}. \qquad (7)$$

In the periodic structure the waves amplitudes are connected also by periodicity condition:

$$U_{2r} = U_{1i}e^{-i\gamma P}, \ U_{2i} = U_{1r}e^{-i\gamma P}, \qquad (8)$$

where $\gamma$ is unknown wave propagation constant in unlimited periodic structure. Here it is more convenient to use parameter $\theta$:

$$\theta = \gamma P, \qquad (9)$$

It describes the phase shift on the period of the structure.

Excluding amplitudes of reflected waves from (6) and using equations (8) one can get an homogeneous system of linear algebraic equations for amplitudes of incident waves:

$$U_{1i}S_{11} + U_{2i}(S_{12} - e^{i\theta}) = 0, \qquad (10)$$
$$U_{1i}(S_{12} - e^{i\theta}) + U_{2i}S_{11} = 0.$$

The condition of equality to zero of determinant of system (10) allows us to calculate the unknown parameter $\theta$:

$$\theta = \pm\arccos\left(\frac{1 - S_{11}^2 + S_{12}^2}{2S_{12}}\right) + 2\pi m. \qquad (11)$$

Note, parameter $\theta=\beta+i\alpha$ is a complex quantity. In order to determine the unique quantity $\theta$ one should correctly choose the branch of arccos function in (11). It could be done using the following conditions. Let's consider that $\beta$ lays within interval:

$$-\pi < \beta < \pi \qquad (12)$$

Then we will be interested in waves that propagate along the positive direction of 0-*z* axis. For propagation direction we mean the direction of energy propagation. From physical considerations it is clear that the wave should decay along the propagation direction. From this condition follows the inequality that provides the wave attenuation:

$$\alpha < 0 \qquad (13)$$

Inequalities (12) and (13) allow choosing the uniquely branch of solution (11) that corresponds to the physically existed wave. Note that real part of *θ* could be both positive and negative. Negative values of *Re θ* correspond to backward waves, while the positive ones - to direct waves.

For backward wave propagation mode [16] the main parameter is the electrical length of transmission line $\theta_0$. A narrow transmission band in the frequency range 2.97 – 2.98 GHz appears within the relatively wide background stop-band for $\theta_0 = 90^\circ$ that corresponds to large values of *α*. Appearance of transmission frequency band in 1D array is connected obviously with the existence of the same band in the single cell. Note, in transmission band *β*>0 and corresponds to the case of direct wave propagation.

Behaviour of parameters *α* and *β* are rather complicated: single transmission band is divided into two parts with a narrow band-stop in between. At low-frequency part of the transmission band α>0, while in high-frequency part α<0. Thus, in the high- frequency part of the transmission band one could observe propagation of backward waves. The bandwidth of backward wave propagation may increase with electric length $\theta_0$. Impedance sharply increases in transmission band, and it is close to zero at stop-band frequencies. Non-zero impedance is caused by finite loss factor.

## 3. Conclusion

Thin film design for meta-atom with Josephson multijunction SQIF-structures was developed. Proposed model topology of 1D meta-media built from series-connected metamaterial atoms based on multijunction superconducting circuit provides propagation of backward waves. Its important feature is the opportunity of resonance frequency tuning of magnetic dipole by means of Josephson inductance control by external magnetic field in SQIF-structure. Taking experimental data for SQIF structure we calculate the bandwidth of backward waves that values 1 MHz, which lays within tuning range of magnetic resonator estimated as 54 MHz.

We thank S. Anlage, I.V. Borisenko, M. Fistul, Y.V. Kislinskii, ,S. Shitov, I.I. Soloviev, A.V. Ustinov, for useful discussions. The work was supported by the Russian Academy of Sciences,

Russian Ministry of Education and Science, grant of the President of Russia Leading scientific schools NSh-2456.2012.2, RFBR grants 12-02-31587mol_a and 12-02-01352-a.

Figure caption

Fig.1 Frequency dependencies of matrix parameters $S_{11}$ (curve 1) и $S_{21}$ (curve 2) calculated for a model of meta-atom. The circuit of meta –atom is presented in the inset.

Fig.2. Thin film topology of the meta-atom on $NdGaO_3$ substrate. 1 and 2 are input/output ports of microwave strip-line, 3 – radiating resonator ("electric dipole"), 4 - spiral superconducting resonator with SQIF ("magnetic dipole")

Fig.3. SQIF design of 30 series-connected SQUID loops with areas in the range $a_i$=50–400 μm$^2$. The width $w$ of the Josephson junctions is 10 μm. The top shows a zoom view of the bottom layer with a part of the SQIF. The equivalent circuit with an SQIF is swhown on the bottom.

Fig.4. Simulated parameters $S_{21}$ and $S_{11}$ for meta-atom, shown in Fig.2. Inset shows 54 MHz change in the resonance frequency when total inductance of Josephson junctions in SQIF changed by 27 pH.

Fig.5. Frequency dependences of $\theta$ parameter. Real part $\beta$ is denoted by curve 1, imaginary $\alpha$ is curve 2. Inset shows the model of 1D meta-media.

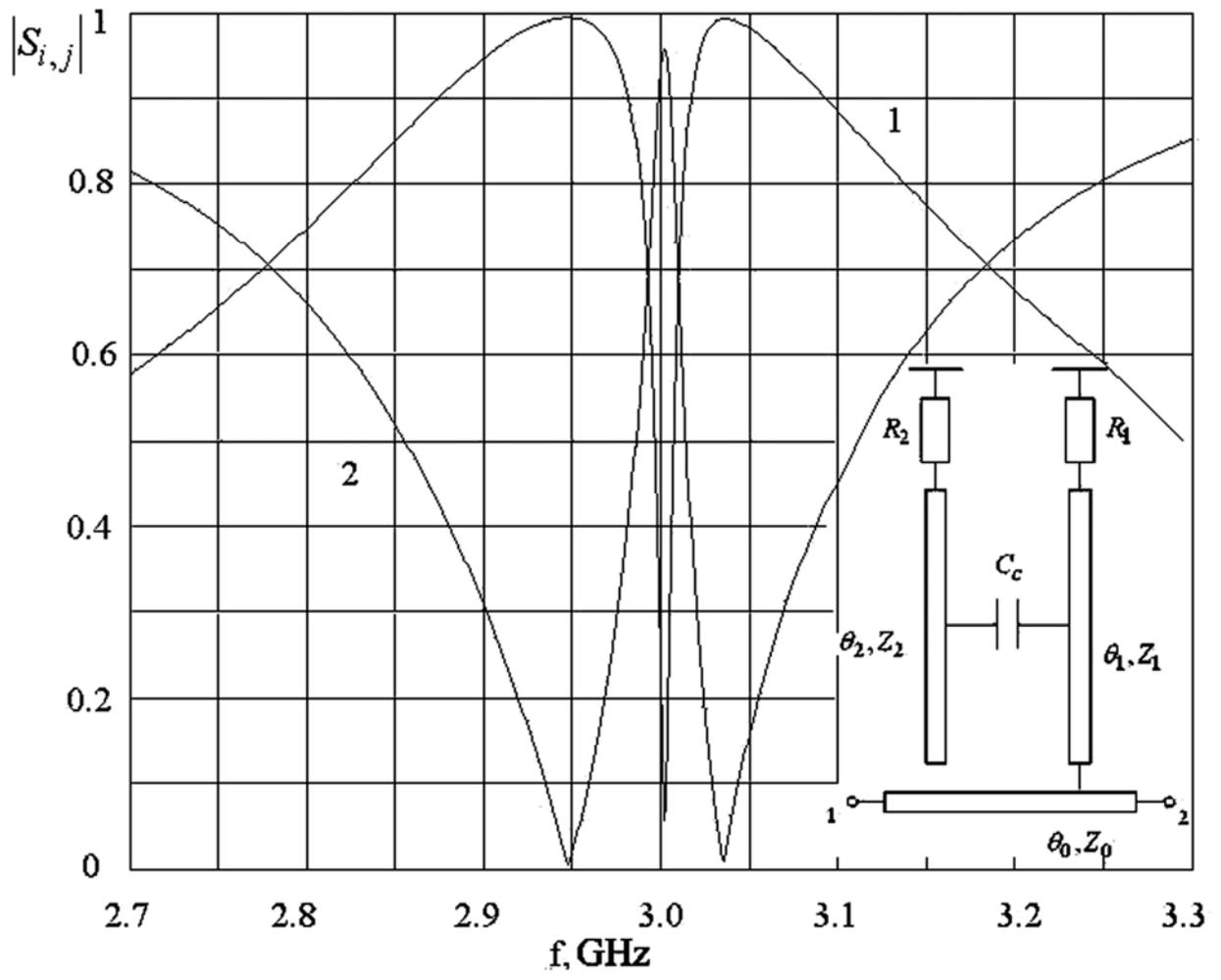

Fig. 1.

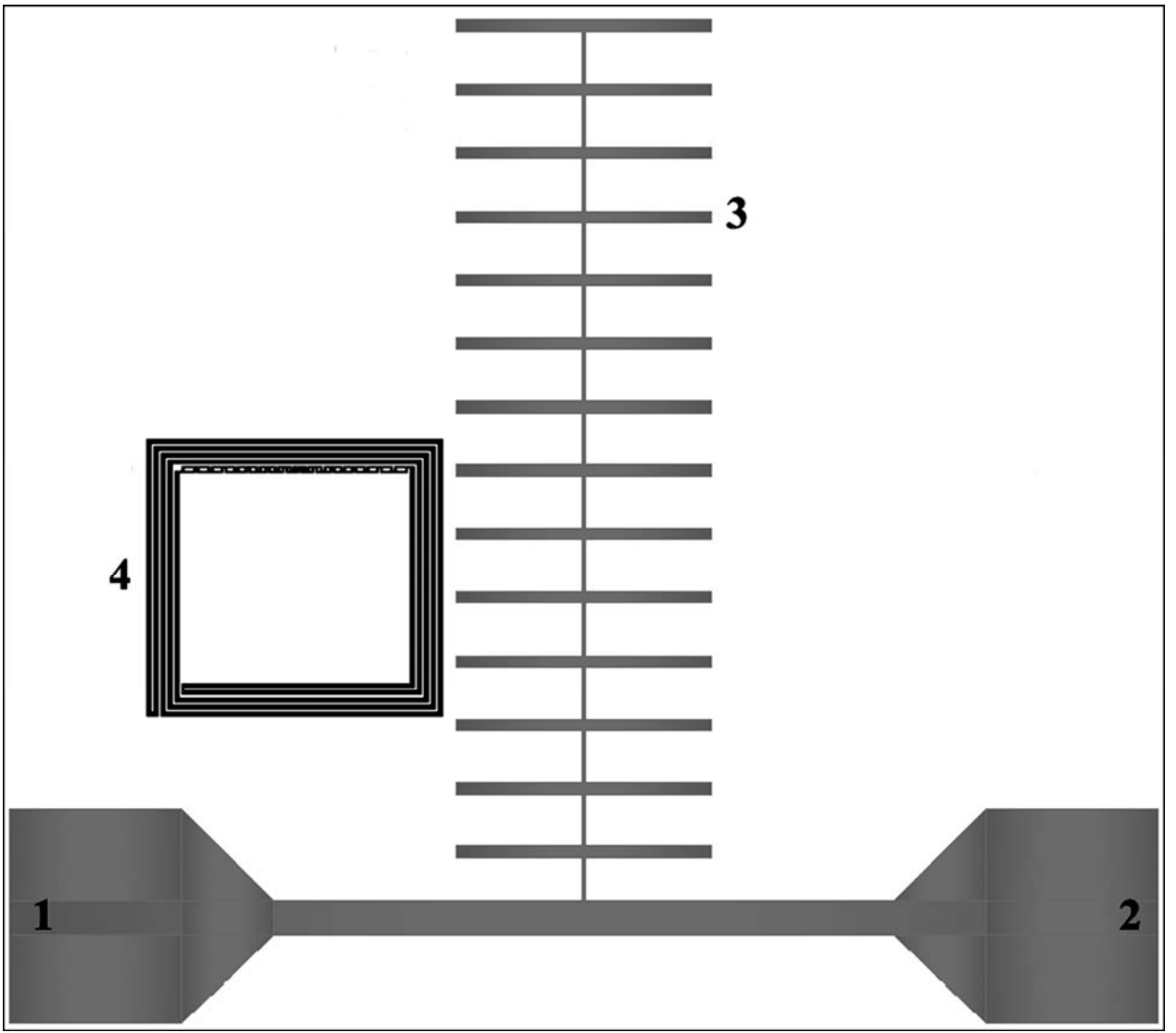

Fig. 2.

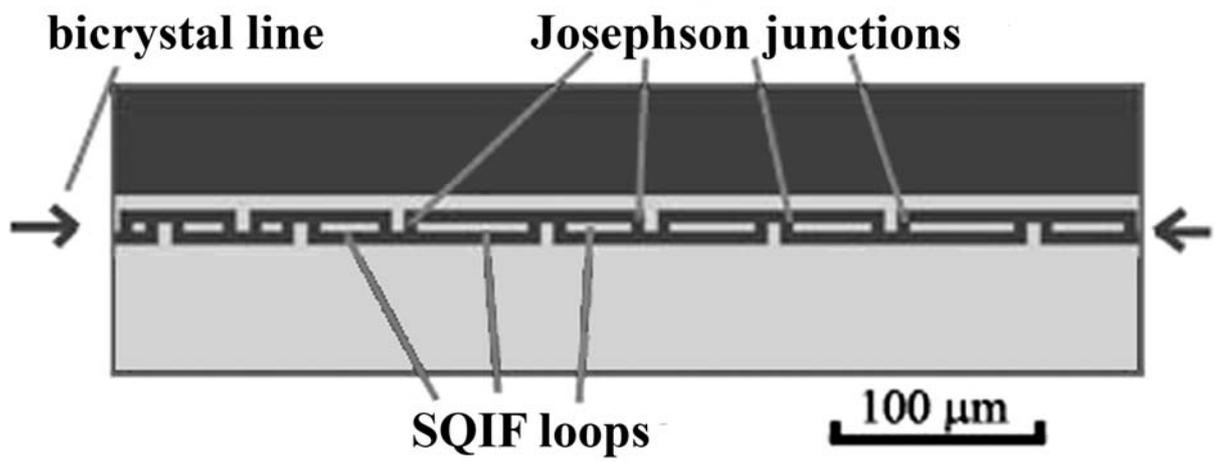
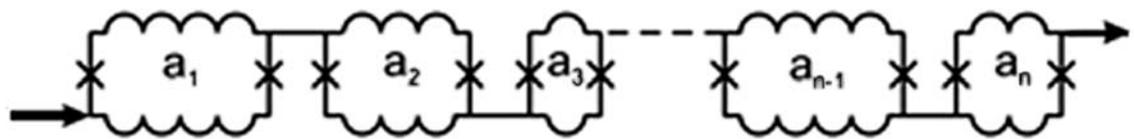

Fig. 3.

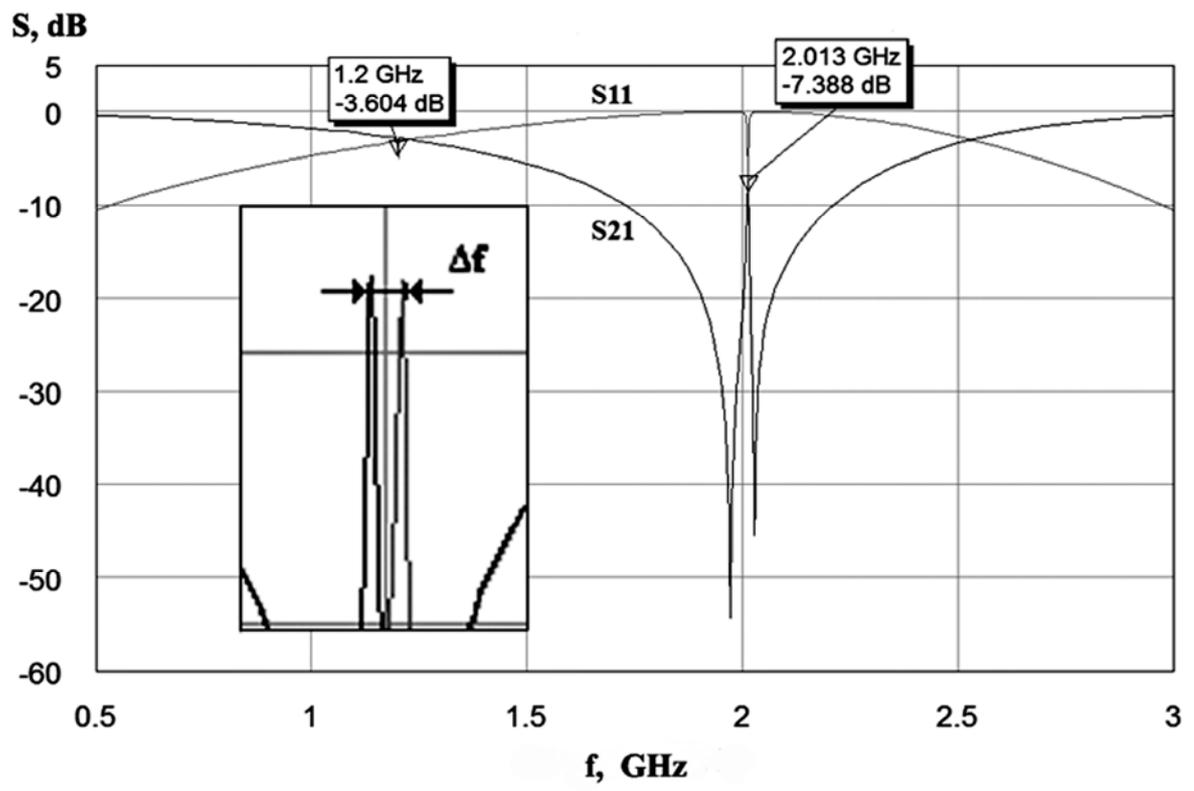

Fig. 4.

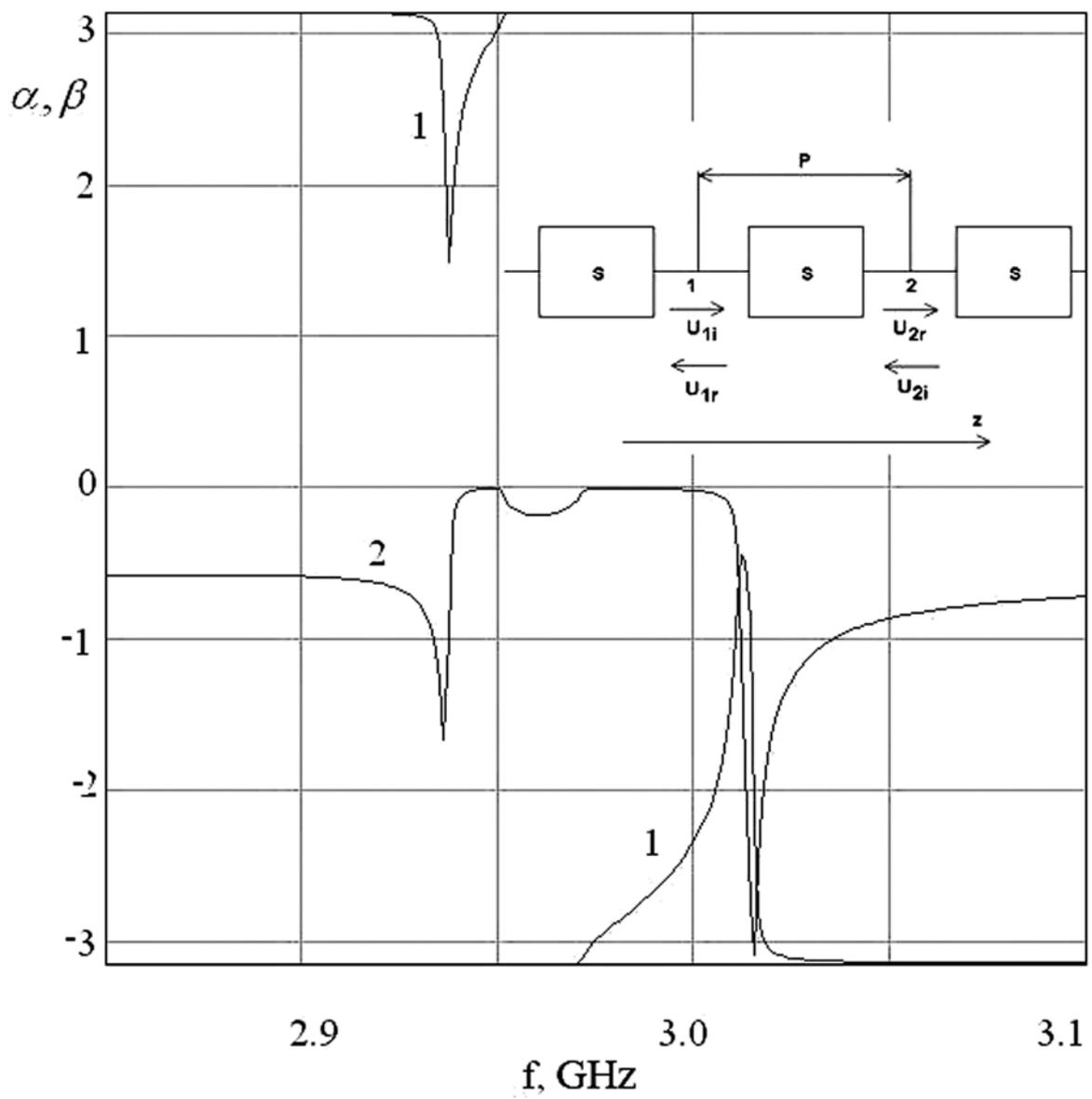

Fig.5